\documentclass{aa}  

\usepackage{graphicx}
\usepackage{txfonts}
\usepackage[normalem]{ulem}
\usepackage[dvipsnames]{xcolor}
\begin{document}

   \title{On the source of the Fe K$\alpha$ emission in T Tauri Stars}

   \subtitle{Radiation induced by relativistic electrons during flares. An application to RY Tau.}

   \author{Ana I. G\'omez de Castro\inst{1,2} 
          \and
          Anna Antonicci\inst{3}
          \and 
          Juan Carlos Vallejo\inst{1,2 }
          }

   \institute{AEGORA Research Group - Joint Center for Ultraviolet Astronomy, Universidad Complutense de Madrid, Plaza de Ciencias 3, 28040 Madrid, Spain
              \email{}
              \and
                Departamento de Fisica de la Tierra y Astrofisica, Fac. de CC. Matematicas, Plaza de Ciencias 3, 28040 Madrid, Spain
                \and 
                I.I.S. Enrico Fermi,  Via Monte Nero, 15/A - 28041 Arona,  Italy
                \\
             \email{aig@ucm.es}
             \thanks{}
             }

   \date{Received ; accepted }

 
  \abstract
   {T Tauri Stars (TTSs) are magnetically active stars that accrete matter from the inner border of the  surrounding accretion disc; plasma gets trapped into the large scale magnetic structures and falls onto the star, heating the surface through the so-called accretion shocks. The X-ray spectra of the TTSs show prominent Fe II~K$\alpha_1,\alpha_2$ fluorescence emission at 6.4~keV (hereafter, Fe~K$\alpha$ emission) that cannot be explained in a pure accretion scenario since its excitation requires significantly more energy than the maximum available through the well constrained free-fall velocity. Neither, it can be produced by the hot coronal plasma.}
   {TTSs display all signs of magnetic activity and magnetic reconnection events are expected to occur frequently. In these events,  electrons 
   may get accelerated to relativistic speeds and their interaction with the environmental matter may result in  Fe~K$\alpha$ emission. It is the aim of this
   work to evaluate the expected Fe~K$\alpha$  emission in the context of the TTS research and compare it with the actual Fe~K$\alpha$ measurements 
   obtained during the flare detected while monitoring  RY Tau  with the XMM-Newton satellite.}
   {The propagation of high-energy electrons in dense gas generates a cascade  of secondary particles that results in an electron shower
   of random nature whose evolution and radiative throughput is simulated in this work using the Monte Carlo code PENELOPE.  A set of conditions representing the environment  of the TTSs where these showers may impinge has been taken into account to generate a grid of models that can aid to the interpretation of the data. }
   {The simulations show that the electron beams produce a hot spot at the point of impact; strong Fe~K$\alpha$  emission and  X-ray continuum radiation are produced by the spot. This emission is compatible with RY Tau observations.}
   {The Fe~K$\alpha$  emission observed  in TTSs could  produced by beams of relativistic electrons accelerated in magnetic reconnection events during
   flares.}


\titlerunning{X-ray excess of T Tauri stars}
\authorrunning{Gomez de Castro et al.}

   \maketitle
%

\section{Introduction}

Low mass pre-main sequence stars  (PMS) are strongly magnetized sources with fields shaped in complex topologies \citep{2019A&A...622A..72V, 2011ApJ...729...83Y, 2009ApJ...704.1721P} that interact with the magnetized outflows launched from the accretion disc \citep{2015MNRAS.450..481D, 2013MNRAS.433.3048K,2012MNRAS.421...63R}.   In the process, matter from the inner border of the  disc gets trapped into large scale magnetic structures and falls onto the star, heating the surface through the so-called accretion shocks.  X-ray emission is expected to be produced at the shock fronts
and contribute to the overall X-ray radiation especially, in heavily accreting TTSs \citep{1998ARep...42..322L,2002ApJ...567..434K,2005A&A...432L..35S}. TTSs  also display an intense flaring activity that shows in the X-ray spectrum \citep{2005ApJS..160..401P,2005ApJS..160..423W,2005ApJS..160..469F,2005ApJS..160..503T}.

The X-ray spectrum of the TTSs  displays some prominent spectral features and, among them, a prominent Fe~II~K$\alpha _1, \alpha _2$ unresolved doublet at 6.4~keV (hereafter, the Fe~K$\alpha$ line), which is observed during the flares and whose origin remains uncertain \citep{2016ApJ...826...84S,2005ApJS..160..503T,2010A&A...520A..38C}. The Fe~K$\alpha$  emission cannot be explained by accretion since the excitation of the line requires significantly more energy than the maximum available through  the well constrained free-fall velocity \citep{2016ApJ...826...84S,2008A&A...481..735G}.  Nor can it be produced by the hot coronal plasma in the stellar atmosphere; this plasma indeed, causes the broad blend of Fe~XXV lines  (6.5~keV-6.7~keV) \citep[see,][for details]{2004ApJ...605..921P} often observed during flares in the Sun and other magnetic stars (including TTSs). 

The excitation of the Fe~K$\alpha$ transition requires cold material, such as that present in the stellar photosphere or at the inner border of the accretion disc, that is excited by X-ray irradiation or by collisions with fast particles, resulting in the subsequent fluorescence emission at Fe~K$\alpha$.  In the flares observed on the Sun and on main-sequence stars
the 6.4 keV line is usually much weaker than the 6.7 keV line as a result, it is expected that the Fe~K$\alpha$ excess of the TTSs is caused by the excitation of disc material.  Previous work has addressed, the relevance of X-ray irradiation from the stellar corona in this process, but requires the irradiation of a large fraction of the disc surface to account for the observed  flux, {\it i.e.} it requires that the disc is flared  \citep{2008ApJ...678..385D}. However, there is not yet and evaluation addressing the relevance of fast particles collisions in the line excitation. This work address this issue.

In this work, we calculate the X-ray radiation produced by the interaction between the beams of relativistic electron (hereafter, e-beams) produced in the flares with the surrounding matter.
Electrons are accelerated to relativistic speeds during magnetic reconnection. In the Sun,  electrons may reach energies in the MeV range during the {\it impulsive phase} due to the onset of plasma micro-instabilities in the reconnection process (see {\it e.g.} \citet{1977ApJ...216..123H}). The  impact of such relativistic e-beams in circumstellar structures, such as the inner border of the disc, or on the stellar surface  generates a cascade of secondary particles that represents an effective degradation of the energy or shower which heats the environment and produces X-ray radiation.
If the medium is dense enough, the propagation of the e-beams results in the generation of a hot spot at the impact point where most of the energy is released into heating. Otherwise, in low density media, the e-beam propagates releasing just a tiny fraction of its kinetic energy into the environment. As a result, the output X-ray spectra from the interaction of e-beams with a given medium depends strongly on the e-beam energy but also on the properties of the medium  \citep[and the view point, see][]{2005A&A...432..443A,2006AN....327..989G}. It is the purpose of this work to compute a network of models that represent the expected X-ray signature of this interaction in the context of  TTSs research. 

This work is presented as follows. In Sect. 2, we describe the code and the general assumptions behind it.  The impact of the  e-beam 
in the surrounding medium results in a random shower of secondary electrons whose interaction with the environment is evaluated using a Monte Carlo code due to its random nature. 
For this purpose, we will use the PENELOPE code which is described in this section. In Sect.3, a first set of simulations
evaluating the output X-ray from the impact in uniform cloudlets is analyzed. In Sect. 4, the output radiation expected from the impact the e-beam on the 
inner border of the accretion disc  described and in  Sect. 5, these results are applied to the interpretation of the X-spectrum of RY Tau during the 2013 superflare
\citep{2016ApJ...826...84S}. The article concludes with a short summary of the main results (Sect. 6).

\section{Numerical simulations and set-up}

The simulations have been carried using the Monte Carlo code PENELOPE which requires the specification of the characteristics of an e-beam and the 
environment through which it propagates. In this section, we describe the PENELOPE code and its tuning to address the propagation
of e-beam in the TTSs environment. We also describe, which are the characteristics of the e-beams considered. We defer 
for the followings sections the description of the two specific grids of simulations run for this investigation.

\subsection{ The MonteCarlo code PENELOPE}

The propagation of a high-energy electrons in dense gas generates a cascade  of secondary particles;
after each interaction, the energy of the primary (or secondary)  particle is reduced and, as a result,  the evolution represents an effective 
degradation of the energy or  shower; the energy is progressively deposited into the medium
and  shared by an increasingly larger number of particles. 

The  evolution of an electron shower is of a random nature thus, 
the process is amenable to Monte Carlo simulation.
 The simulation of photon transport is straightforward since the mean 
number of events in each history is small. In practice, high
energy photons are absorbed after a single photoelectric 
or pair-production interaction or after a few Compton interactions. 
The simulation of electron transport is much more difficult 
since the average energy loss by an electron in a 
single interaction is very small (around a few tens of eV). 
As a consequence, high-energy  electrons suffer a large number of 
interactions before being effectively absorbed in the medium. Hence, 
detailed simulation is feasible only when the average number of collisions 
in the path is not too large. 

For high-energy electrons, most of the Monte Carlo codes rely 
on multiple scattering theories which 
allow the simulation of the global effect of a large number of events 
in a track segment of a given length (or {\it step}). This 
procedure is referred  at  as ``condensed'' Monte Carlo method since 
the global effect of a large number of events is condensed in a 
single {\it step}. 

In this work, we use the program developed by \citet{Antonicci2004} to simulate
the propagation of fast electrons in plasmas. This program runs 
a subroutine package  named PENELOPE 
which is a Monte Carlo code\footnote{The Monte Carlo algorithm 
implemented in PENELOPE incorporates a ``mixed'' scattering model 
for the simulation of electron and positron
transport. Hard interactions, with scattering angle and/or energy loss
greater than preselected cutoff values are simulated in detail, by using
simple analytical differential cross sections for the different
interaction mechanisms and exact sampling methods. Soft interactions, with
scattering angle or energy loss less than the corresponding cutoffs, are
described by means of a multiple scattering approach. The 
code is user-friendly  and incorporates photon cross-section data from the EPDL97
which includes new libraries for the low-energy  photon cross-sections, 
such as XCOM and EPDL97. The code is available at the web site
of the International Atomic Energy Agency 
(\url{http://www.nea.fr/lists/penelope.html})}
initially designed to study the 
propagation of electron-photon showers, with energy from 
100~keV to several hundred MeV, in arbitrary materials, see \citet{Baro1994}  and references
therein. The cross sections for hard elastic scattering,
hard inelastic collisions, hard  bremsstrahlung emission,
soft {\it artificial} events and positron annihilation
are taken into account to calculate the interactions
of the electrons with matter. 

The following cross sections
have been taken into account for the interaction of the
secondary photons with the cloud:
coherent (Rayleigh) scattering,
incoherent (Compton) scattering, photoelectric absorption
of photons and electron-positron pair production.

We have set a minimum energy threshold
of 1~keV for the electrons since electrons with lower energy 
are not able to induce significant radiative effects in the X-ray
range in the gas. The program  is described in full detail in \citet{Antonicci2004} . It
has also been  adapted to simulate 
fast electron propagation in solid matter within the 
context of laser-matter interaction. This
has allowed validating the collisional part of the code 
by comparing the computational results with the experimental 
results from  LULI (Laboratoire pour l'utilisation de lasers intenses) 
and from the Livermore laboratory\footnote{The code is, in addition,
widely used in the medical community and, for instance,
\citet{Sung2004}  have recently verified it for clinical 
dosimetry of high-energy (10~keV-150~keV) electron and photon beams.}.

\subsection{The properties of the e-beam}

In the context of the TTS research, the most luminous e-beams are accelerated in the reconnection events associated with strong flares. 
The  X-ray light curves  display a rich phenomenology of events; from 
compact flares similar to those observed in magnetically active stars,  to extended flares proposed to be associated with large-scale loops
that may reach about few stellar radii in length \citep[see i.e.][]{2005ApJS..160..469F} and would require average field strengths at the base 
of the corona of  $\sim 100 - 150$ kG \citep{2006MNRAS.367..917J}; the density of the plasma within these loops
is  $ 10^{10} -  10^{11}$~cm$^{-3}$.

A basic estimate of the expected mechanical luminosity of these strong e-beams can be obtained from the power 
needed to  accelerate the particles ($L$), which is roughly given by,
\begin{equation}
\begin{tabular}{ll}
$L $ & $=\frac{ B^2}{8\pi} V_A \delta A $  \\
&$= 0.85 \times 10^{30}  \left( \frac{erg}{s} \right) \left( \frac {B}{100 G} \right)^3 \left( \frac{n}{10^{10} cm^{-3}}\right) ^{-0.5}
 \left( \frac {\delta A}{10^{19} cm^2} \right)$\\
 \end{tabular}
\end{equation}
\noindent
where B, is the magnetic field, $\delta A$ , the section of the reconnecting current and, $V_A$, the Alfv\'en velocity of the plasma given by, 
\begin{equation}
V_A = \frac {B} { (4\pi m_H n)^2}
\end{equation}
\noindent
and $m_H$, the mass of the Hydrogen atom and $n$, the particle density. The most uncertain parameter in this calculation is the surface of the 
reconnecting layer. For instance, for the super-flares described above, it seems reasonable to assume that  magnetic reconnection occurs
when the giant magnetic loop meets the inner border of the disc. Given the typical radius and the expected scale-height of the inner border of the disc 
(see Sect. 4), the surface of the re-connecting layer could be as large as $\sim 10^{19}$~cm$^2$ and luminosity of the flare
as high as  $2 \times 10^{-4} L_{\sun}$ (see Eq.~1). In practice, only a fraction of the magnetic energy released in the flare goes
into particles acceleration; according to recent observations of Solar flares, this fraction is about $\sim 50$\% of the total energy
 \citep{2019ApJ...872..204B}. 

However, this bulk  luminosity is significantly smaller than the rise of the X-ray flux observed during flares in TTSs. 
For instance, the X-ray luminosities reach on average $<L_{X}> = 10^{32}$~erg~s$^{-1}$ during strong flares but some events 
get as strong as  $L_{X} = 0.25 L_{\odot}$; see {\it i.e.}  the event detected in AN Ori during the Chandra Orion Ultradeep Project (COUP) survey  \citep{2005ApJS..160..469F} or the flare detected with the ASCA satellite during the observation of V773~Tau \citep{1998ApJ...503..894T}. 
Thus,  given the uncertainties involved, we have opted to compute the expected X-ray spectrum for a set of  
e-beams with  mechanical luminosities ranging from $2 \times 10^{-4} L_{\sun}$ up to $2 \times 10^{-1} L_{\sun}$.

The energy distribution of the electrons in the beam can be modelled by a power law \citep[see][for an in-depth discussion]{2008PhRvL.101t5004O}. 
In particular, the distribution function of the number of electrons ($n_e$) per kinetic energy ($\epsilon$) is often described, as a single power 
law with index -4. For our application,  the  distribution could be truncated to energies between 30 keV and 2 MeV:
 
\begin{equation}
n(\epsilon) \propto \epsilon ^{-4}, \hspace{0.5cm} {\rm for }    \hspace{0.1cm} 30 {\rm keV} \leq \epsilon \leq 10 {\rm MeV}
\end{equation} 
\noindent

The low energy cut-off is set to disregard the contribution of low energy electrons since they produce optically thin Bremsstrahlung radiation
and no Fe~K$\alpha$ emission.
The upper threshold is set to optimize the computation time of the simulations. This limit guarantees reaching an accuracy of $10^{-4}$ in the total energy, which is 
more than enough for the the accuracy of the X-ray observations of the TTSs (and the purpose of this work). 

Using a power law distribution presents however, some drawbacks for the diagnoses intended in this work.  In particular,
the mean kinetic energy of the electrons ($<K>$) cannot be tuned to evaluate to what extent is significant in the output Fe~K$\alpha$ emission.
For this reason, we have opted for a Maxwell relativistic (MR) distribution  and worked with e-beams of different mean kinetic energies.
The mathematical form of the distribution function of  the number of electrons ($n_e$)  with the kinetic energy ($K$) and is given by,
\begin{small}
\begin{equation}
\frac{dn_e}{dK} = \gamma (\gamma ^2 -1)^{1/2} e^{-K/k_BT_e}
\end{equation}
\end{small}
\noindent
with $\gamma = (1-(v/c)^2)^{-1/2}$ and $K = 0.511 MeV(\gamma -1)$; $k_B$ is the Boltzmann constant and $T_e$, the electron temperature.
Note that the mean kinetic temperature of the e-beam is $\sim 3 k_B T_e$ since the distribution has a broad tail toward high energies. 

Finally, the e-beam is assumed to have cylindrical symmetry and the electrons be distributed within any given section following
a radial distribution such that,
\begin{equation}
n_e (r) = n_{e,0} \exp{\left(-\left(\frac{r}{r_0}\right)^2\right)}
\end{equation}
\noindent
with  $r_0=10^5$ cm. The value of $n_{e,0}$ is set to satisfy the injection of the total mechanical luminosity 
in 100~s.

\section{Basic simulations: propagation through an homogeneous medium}

\begin{figure}
\begin{center}
\includegraphics[width=7cm]{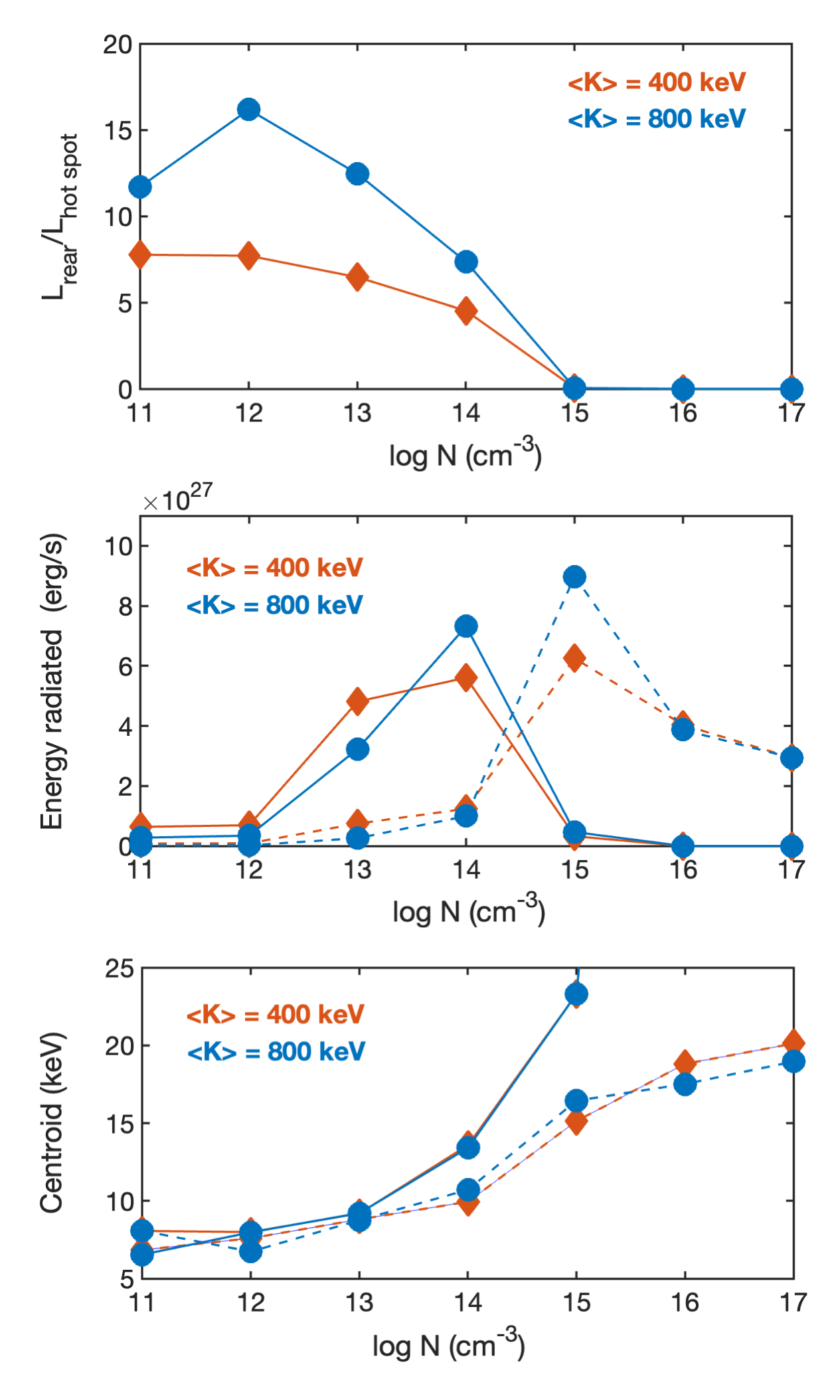}
\end{center}
\caption{General properties of the X-ray spectrum for beam K= 400 keV (red) and K = 800 keV (blue).
Top: ratio between the luminosity radiated from the rear of the cloud and the hot spot, at the front, as a function of the 
density of the cloud. Middle: Luminosity of the spectra as a function of the density. Bottom: dependence of the centroid
of the X-ray energy distribution on the cloud density. The continuous and dashed lines refer to data from the transmitted
and back-scattered spectra. The kinetic energy of the e-beam is fully absorbed within
the cloud for particle densities above $10^{14}$~cm$^{-3}$.   }
\label{}
\end{figure}

\begin{figure}
\includegraphics[width=11cm]{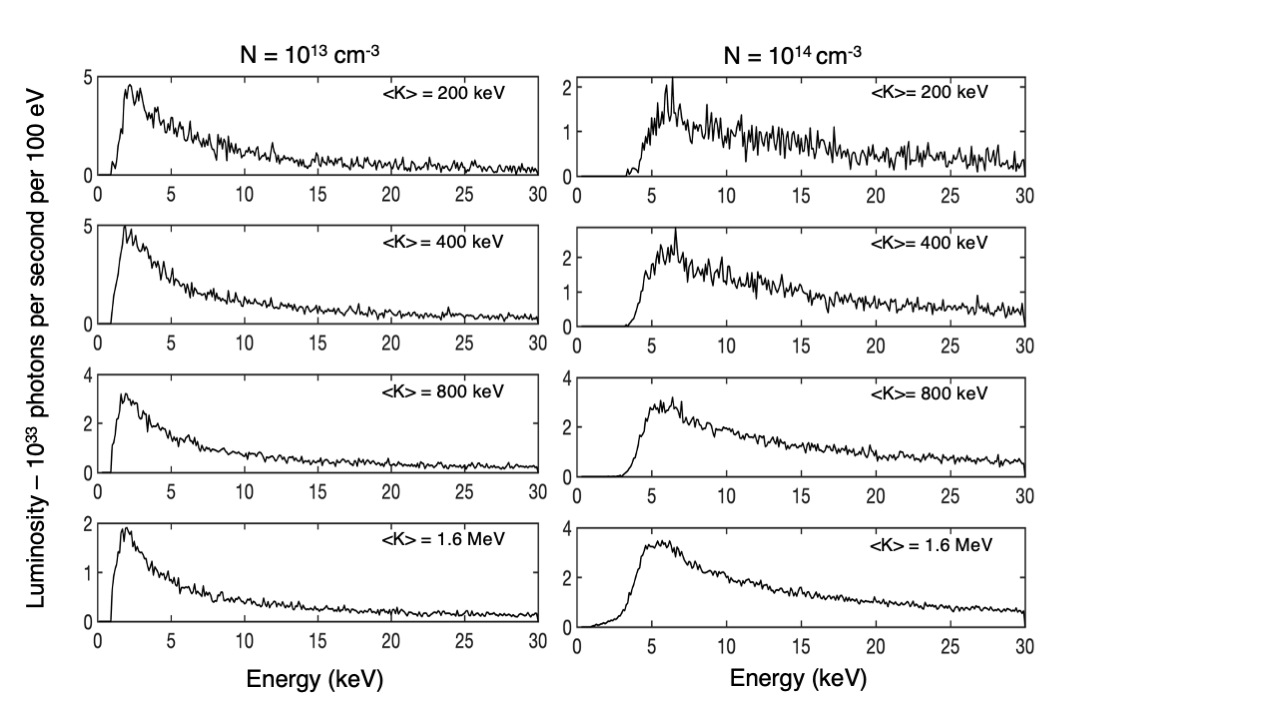}
\caption{Output X-ray spectrum from the rear of the cloud.  }
\label{}
\end{figure}

This grid of simulations has been run to characterize the radiative output produced at X-ray wavelengths (2-30 keV) by the propagation 
of an e-beam through a uniform density medium. 
The medium includes the most abundant elements up to Iron in solar abundance \citep{1982GeCoA..46.2363A}.
In this set-up, the e-beam impinges perpendicularly to the surface of the medium which is a spherical cloudlet of uniform
density and diameter  $10^{10}$~cm.  The simulations have been run for cloudlet densities ranging from 10$^{9}$~cm$^{-3}$  to 10$^{17}$~cm$^{-3}$ 
in powers of ten. The lower limit is set by the density of the accreting matter trapped in the funnel flows that channel the gas from the disc onto the stellar surface; ultraviolet observations of the semi-forbidden C~II]  emission at 2323~\AA\  indicate that the density ranges from $10^9$~cm$^{-3}$ to10$^{12}$~cm$^{-3}$ in these structures \citep{2014MNRAS.442.2951L}. The upper limit is set for the expected mid-plane density at the innermost border of an accretion disk around a solar mass TTS undergoing an accretion rate of 10$^{-6}$~$M_{\odot} yr^{-1}$ (see Sect 2.5).

The mechanical luminosity of the e-beam is 0.002 L$_{\odot}$ and  $<K>$ values
of 200~keV, 400~keV, 800~keV and 1.6~MeV have been tested. In all cases, an MR energy distribution has been assumed
for the electrons in the e-beam.

The e-beam propagation results in the production of X-ray radiation. We have measured the backscattered and forward scattered radiation
from the cloudlet along the direction of propagation of the e-beam. A hot spot is always produced at the entry point however, 
to observe X-ray radiation from the back, the density of the medium needs to be $\leq 10^{14}$~cm$^{-3}$. 
Note that X-ray radiation is fully absorbed for column densities  above $10^{24}$~cm$^{-2}$ \citep{1983ApJ...270..119M} . The characteristics of the output X-ray spectra are summarized in Fig.~1 and  some sample spectra  are shown in Fig.~2 and Fig.~3 for the rear of the cloudlet and the back-scattered radiation from the hot spot, respectively.  In general, the spectra are harder, {\it i.e.}, the centroid of the X-ray energy distribution is observed at higher energies, as the density of the cloud increases. Increasing the mean kinetic energy of the e-beam has the same effect.

\begin{figure*}
\includegraphics[width=19cm]{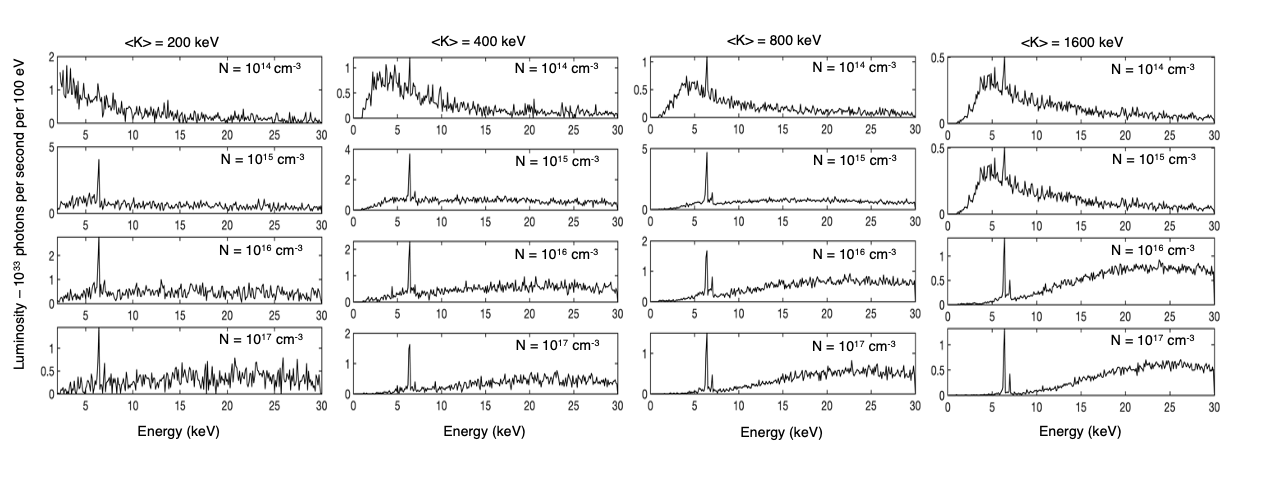}
\caption{Output X-ray spectrum from the hot spot obtained for  kinetic energies of the e-beam: 200 keV, 400 keV,
800 keV, 1.6 MeV (see top) and media density: 10$^{14}$ cm$^{-3}$,  10$^{15}$ cm$^{-3}$, 10$^{16}$ cm$^{-3}$, and,
10$^{17}$ cm$^{-3}$.  No hot spot is produced for the lowest simulated density $N = 10^{13}$ cm$^{-3}$, neither Fe~K$\alpha$ emission.  }
\label{}
\end{figure*}

The spectrum of the hot spot contains some additional interesting features including some prominent emission lines. 
The Fe~K$\alpha$ line (6.4 keV) is observed for $N > 10^{13}$~cm$^{-3}$, as a strong feature in the spectrum;
this line is actually, an unresolved doublet produced in the transitions  $[1s 2S^{1/2}]-[2p 2P^{1/2,3/2}]$ of Fe~II. 
For e-beams with kinetic energy, $K \geq 800$~MeV and  $N \geq 10^{16}$~cm$^{-3}$,  also the Fe~K$\beta$ line 
at 7.1~keV is observed.  As shown in  Fig. 4, the strength of the Fe K$\alpha$ line decreases as the density of the cloud 
increases while the  strength of the Fe~II~K$_{\beta}$ line depends on the energy of the e-beam; also, the ratio 
Fe K$_{\alpha}$/Fe K$_{\beta}$  decreases with the energy of the beam. The broad iron feature at $\sim 6.7$~keV, 
which is usually observed in TTSs, only shows in few spectra (see Fig. 5);  this feature arises from very hot plasma 
in the hot spot at the point of impact of the e-beam.  This Fe XXV feature is probably a blend of the x, y, and z lines 
of Fe XXV  \citep[see][for a detailed description of the main transitions]{2004ApJ...605..921P}.

\begin{figure}[h!]
\includegraphics[width=8cm]{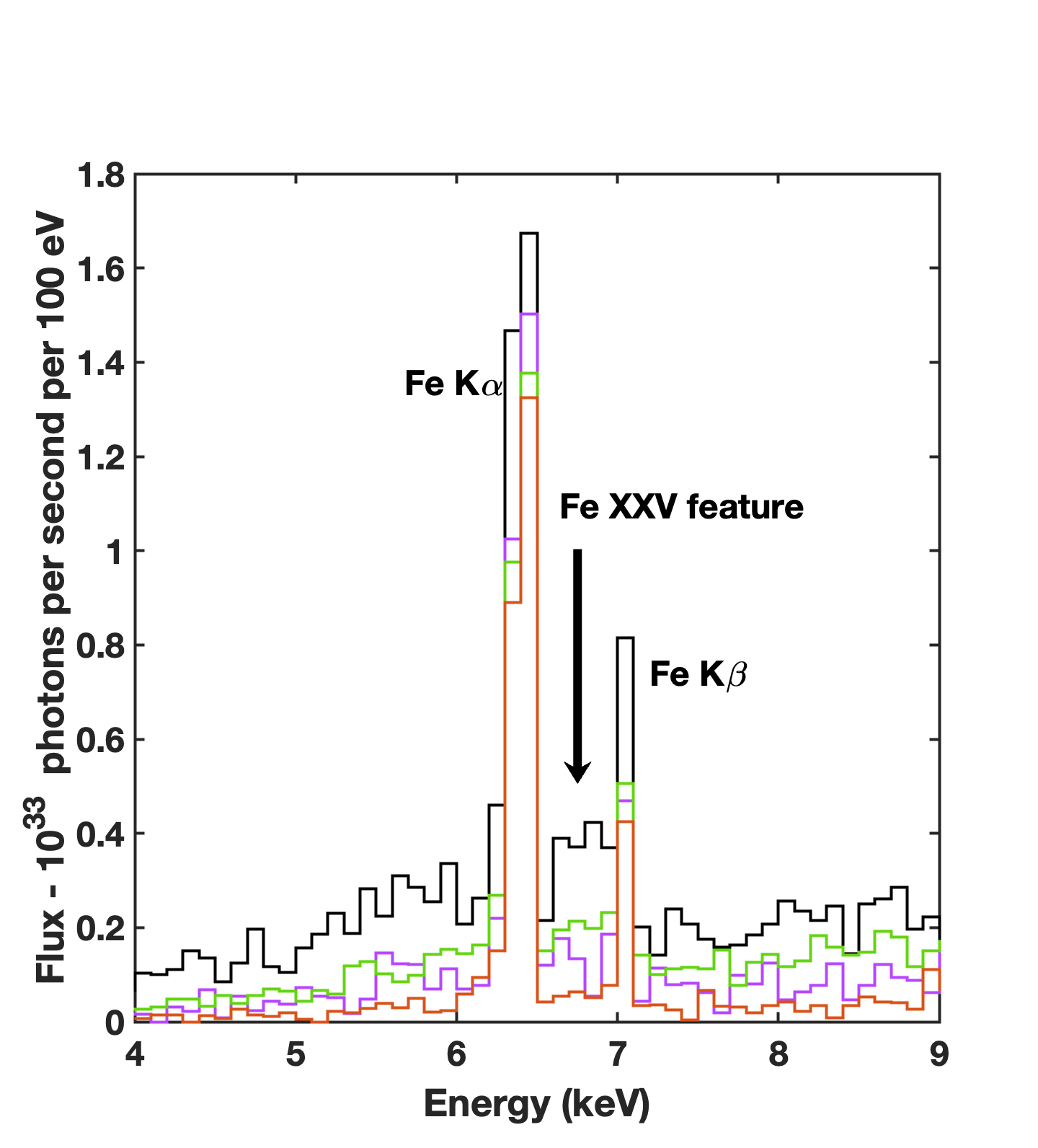}
\caption{The Fe feature in the spectrum of the hot spot (see Fig. 3). Only some relevant simulations are shown:
($\rho = 10^{16}$~cm~s$^{-3}$, $K= 800$ keV) is marked in black,  ($\rho = 10^{17}$~cm~s$^{-3}$, $K= 800$ keV) is marked in purple,
($\rho = 10^{16}$~cm~s$^{-3}$, $K= 1,600$ keV) is marked in green and ($\rho = 10^{17}$~cm~s$^{-3}$, $K= 1,600$ keV) is marked in 
brown.}
\label{}
\end{figure}

To summarize, the main features derived from this grid of simulations are:
\begin{itemize}
\item Fe~K$\alpha$ (and K$\beta$) emission is only observed in the spectra of the hot spot and for densities above 10$^{14}$~cm$^{-3}$.
\item $<K> = 400$ keV  maximizes the Fe~K$\alpha$ emission (see Fig. 5).
\item  Increasing $<K>$ or increasing the density of the cloudlet shifts the centroid, {\it i.e.} makes the output X-ray spectrum harder.
\end{itemize}
In general, this process is inefficient in terms of the X-ray radiation production compared with the input of mechanical energy of the e-beam.
As an example,  in the simulation with $<K> = 400$~keV and $N = 10^{14}$~cm$^{-3}$, the hot spot radiates (in the 
2 - 30 keV) range only $\sim 0.07$\% of the input mechanical luminosity of the e-beam and the radiation from the rear of the cloudlet is ten times 
smaller. Most of the energy is diffuse away of the cloudlet by the scattered electrons and thus, only large dense structures are efficient
in the transference of the e-beam energy into X-ray radiation.

\begin{figure}
\begin{centering}
\includegraphics[width=10cm]{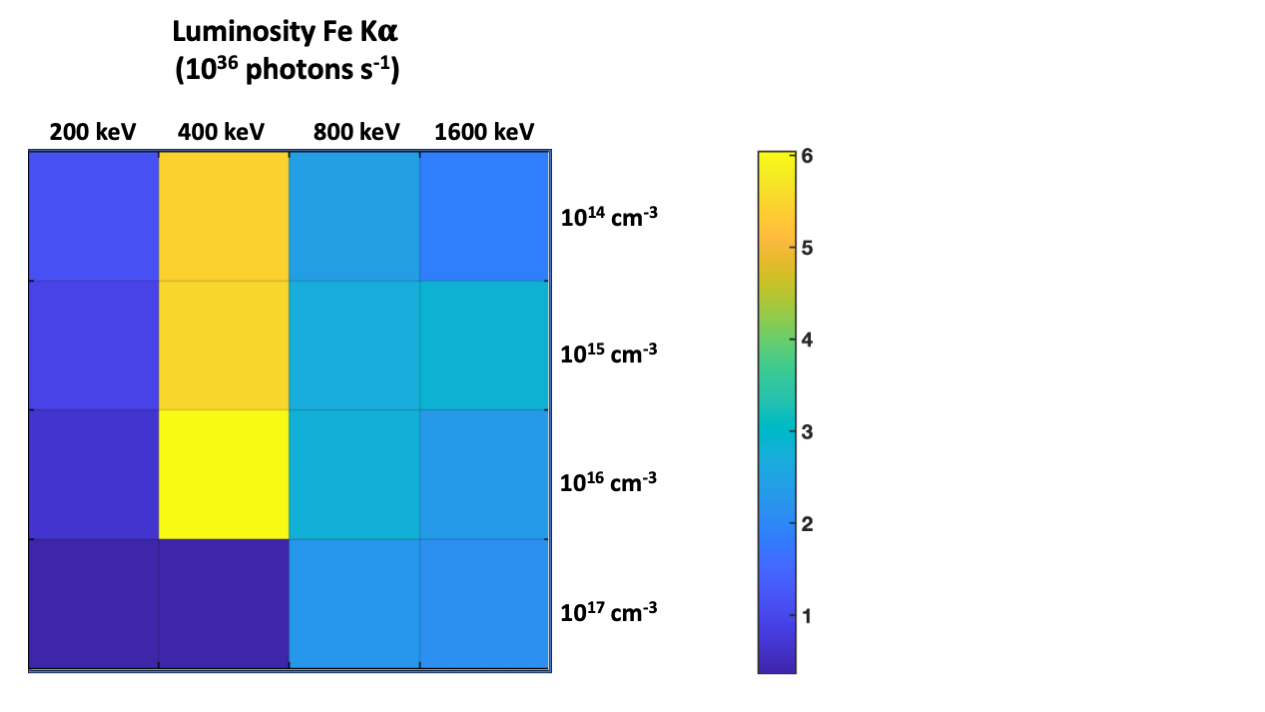}\\
\includegraphics[width=10cm]{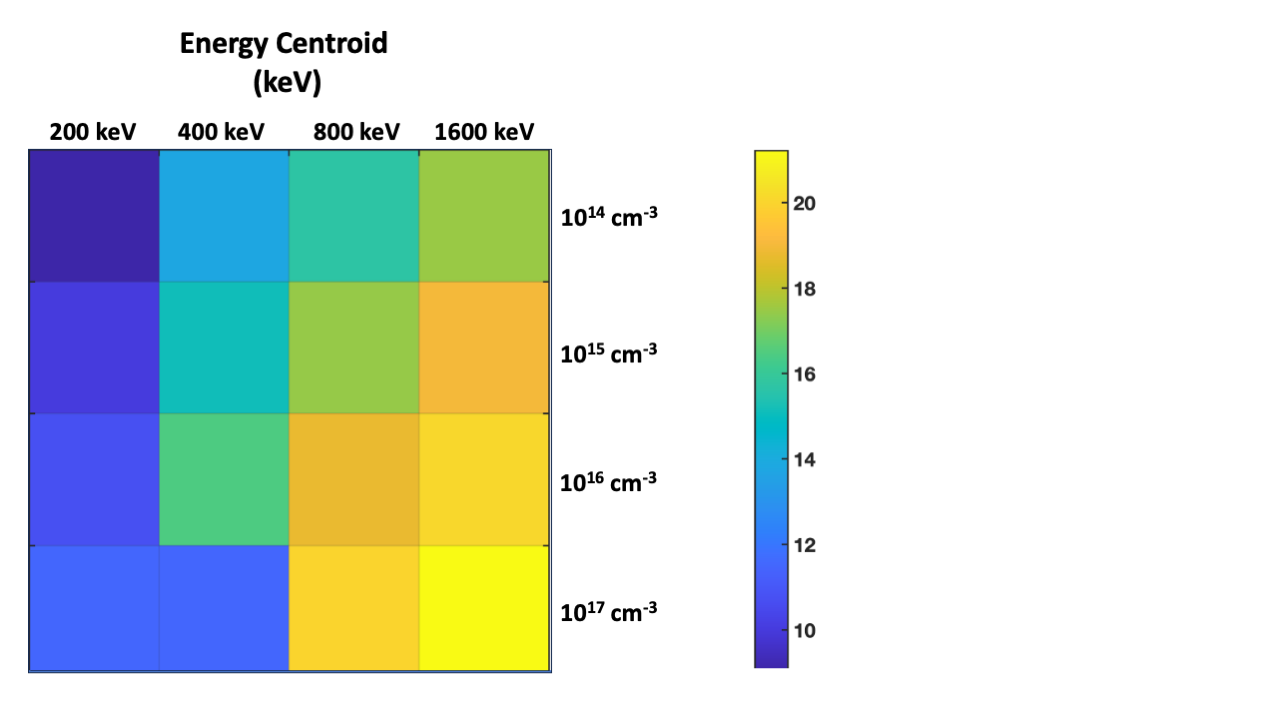}\\
\caption{Variation of the Fe K$\alpha$ flux  (top) and the energy centroid (bottom) of the X-ray spectrum with the cloud density and the mean kinetic 
energy of the beam $<K$.}
\end{centering}
\label{}
\end{figure}

\section{ X-ray output from the impact of e-beams on the accretion disc}

In this set-up, the e-beam is assumed to impinge perpendicularly to the surface of the disk, which is stratified. 
The disc is modelled as a set of plane parallel layers of increasing density to the mid-plane following the vertical structure of a standard  $\alpha$ disc  \citep{1973A&A....24..337S}; the scalings have been adopted from \citet{2013pss4.book..279G}. 
The e-ebeam impacts on the disc from the $z$ axis, as shown in Fig. 6, and  the disc density depends on z as,
\begin{small}
\begin{equation}
\rho (R,z) = \rho (R, 0) exp (- \frac{z^2}{2H^2})
\end{equation}
\end{small}
\noindent
with,
\begin{small}
\begin{equation}
H(R) = 3.4 \times 10^9 cm \left( \frac{\dot M}{10^{-8} M_{\odot} yr^{-1}}\right) ^{1/8}.   \left( \frac{M}{M_{\odot}} \right) ^{-3/8}
\left( \frac{R}{0.014 AU} \right) ^{9/8}
\end{equation}
\end{small}
\noindent
and, 
\begin{small}
\begin{equation}
\rho (R) = 0.94 \times 10^{-8} g cm^{-3}   \left( \frac{\dot M}{10^{-8} M_{\odot} yr^{-1}}\right) ^{5/8}  \left( \frac{M_*}{0.5 M_{\odot}}\right) ^{5/8}
\left( \frac{R}{0.014 AU}\right) ^{-15/8}
\end{equation}
\end{small}
The disc temperature is given by,
\begin{small}
\begin{equation}
T (R) = 1,765 K  \left( \frac{\dot M}{10^{-8} M_{\odot} yr^{-1}}\right )^{1/4}   \left( \frac{M_*}{0.5 M_{\odot}}\right) ^{1/4}
\left( \frac{R}{0.014 AU}\right) ^{-3/4}
\end{equation}
\end{small}
The PENELOPE numerical code is not designed to work with a density profile along the z-axis of the irradiated sample. To overcome this problem, 
the disc has been implemented as set of plane parallel layers of uniform density. The exponential profile described in Eq 5 is reproduced by 10 layers, as shown in Fig. 6 and the mean density changes by a $\sim$25\% between layers.

\begin{figure}
\includegraphics[width=9cm]{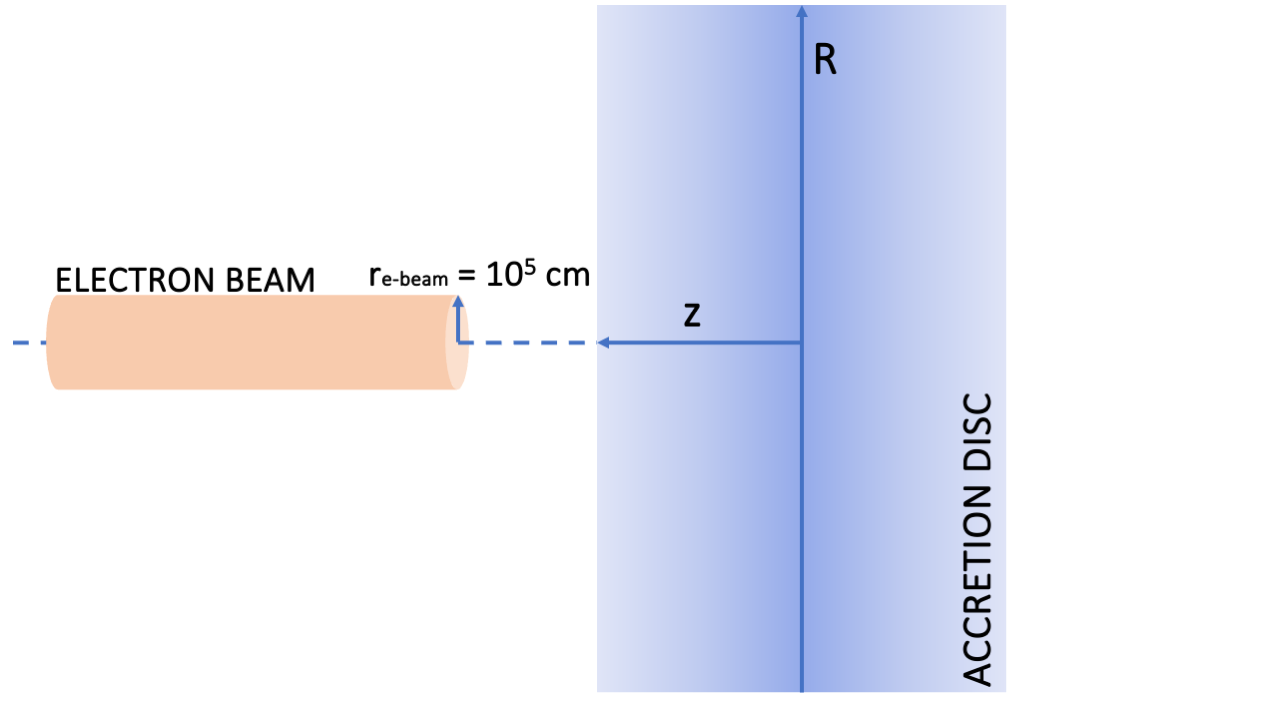}
\includegraphics[width=10cm]{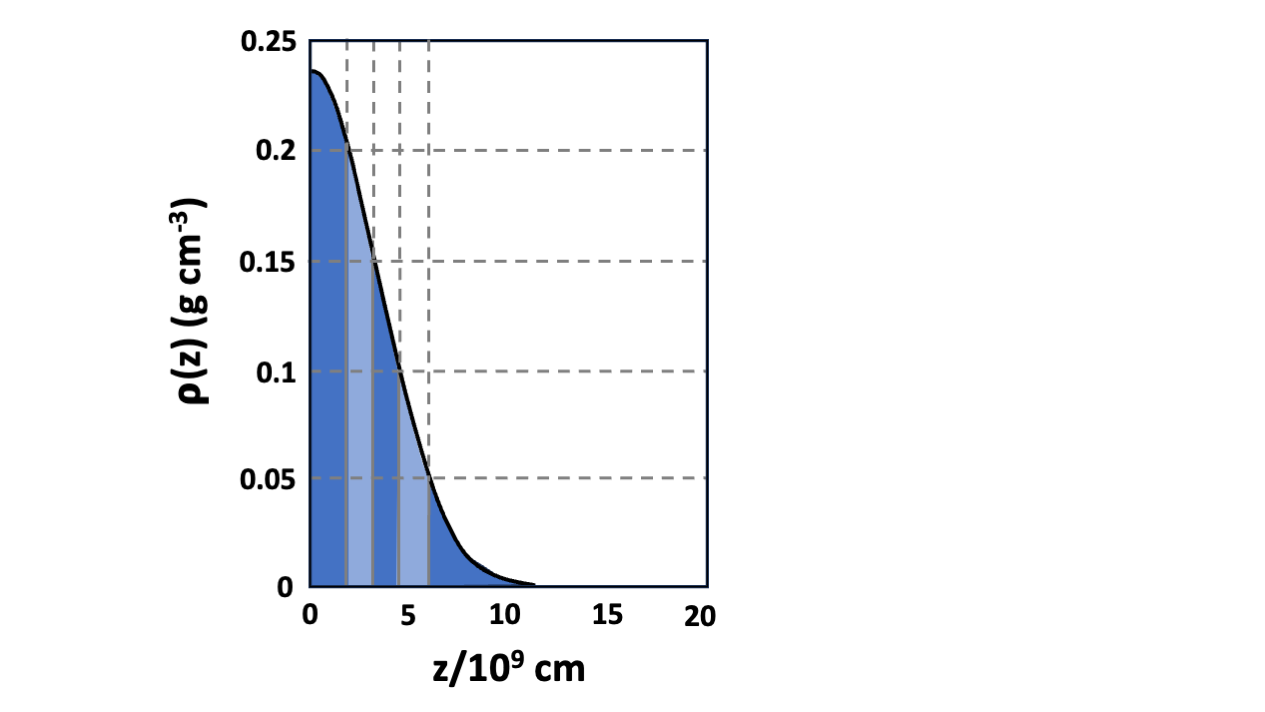}
\caption{Top: Simulation scheme. The e-beam impinges normally on the accretion disc surface. Bottom: The normalized density profile (see Eq. 6) is represented for the inner border of a sample disc with accretion rate $10^{-8} M_{\odot} yr^{-1}$. The structure is symmetric around the disc mid-plane and only the upper part is represented. The density profile from Eq. 6 is plotted with a blue line and the limits of the five layers implemented in the simulation are marked with dashed lines. }
\label{}
\end{figure}

The grid of simulations is summarized in Table 1. As in Sect.~3, the radius of the e-beam is assumed to be 10$^5$~cm and 
the disc composition is set to solar abundances.  The output spectra obtained for an e-beam with $<K> = 800$~keV
are represented in Figs.7 (transmitted) and 8 (hot spot). 
For the range of mechanical luminosities of the e-beam studied in this work ($\sim 10^{-1} - \sim 10^{-4}$~L$_{\odot}$), the X-ray luminosity increases
linearly with the e-beam mechanical luminosity thus, only spectra corresponding to a sample luminosity ($2\times 10^{-3}$ L$_{\odot}$) are represented in the figures.

\begin{table}
\caption{Network of simulations of the interaction of the e-beams with the inner disc $^{*}$}           
\label{table:2}    
\begin{centering}    
\begin{tabular}{l c c c c }      
\hline\hline                 
Simulation & \multicolumn{4}{c}{Properties of the disc$^{1}$} \\

Id. & \cline{1-4} & \\
 
 & Acc. Rate & $H_0$ & $T_0$ & $\rho _0$   \\
		& ($M_{\odot} yr^{-1}$) & (cm)  & (K)  & ($g cm^{-3}$) \\
\hline   
AD-Ma6 & $10^{-6}$ & $6.1 \times 10^9$ & 5,581 & $1.7\times 10^{-7}$ \\
AD-Ma7 & $10^{-7}$ & $4.5 \times 10^9$ & 3,139 & $ 3.9\times 10^{-8}$ \\
AD-Ma8 & $10^{-8}$ & $3.4 \times 10^9$ &1,765 & $ 9.4\times 10^{-9}$ \\
AD-Ma9 & $10^{-9}$ & $2.5 \times 10^9$ &993 & $ 2.2\times 10^{-9}$ \\
AD-Ma10 & $10^{-10}$ & $1.9 \times 10^9$ &558 & $ 5.3\times 10^{-10}$ \\
AD-Ma11 & $10^{-11}$ & $1.4 \times 10^9$ &314 & $ 1.2\times 10^{-10}$ \\
\hline                                   
\end{tabular}

\begin{tabular}{ll}
$^1$ & $H_0$, $T_0$, $\rho _0$ are computed for the mid-plane at $R_0$ for the  \\
& given accretion rate (Eq. 7-9). \\
\end{tabular}
\end{centering}  
\end{table}

As expected, the e-beam is absorbed in high density environments; this translates into accretion rates 
$M_{a} > 10^{-10}$~M$_{\sun}$yr$^{-1}$  in the accretion disc scenario. The transmitted  spectrum is only observed for the most rarified 
accretion discs  and the cut-off energy shifts to higher energies as the accretion rate increases (see Fig. 7). Again, no spectral features are 
observed in the transmitted spectrum.

\begin{figure}
\includegraphics[width=11cm]{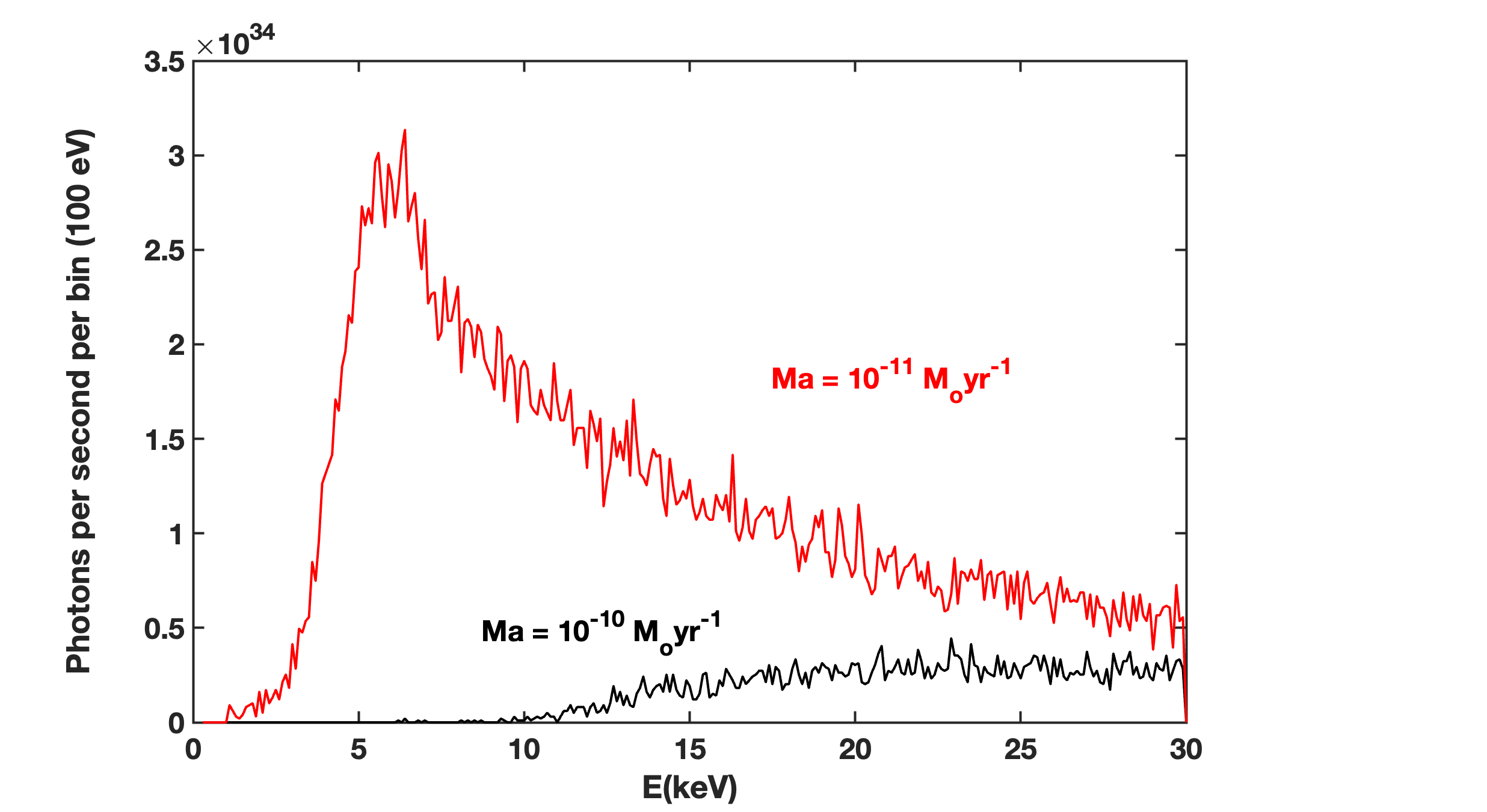}
\caption{Transmitted X-ray spectrum. Simulation with e-beam luminosity: $2\times 10^{-3}$ L$_{\odot}$.}
\label{}
\end{figure}

The  spectrum of the hot spot displays a prominent Fe~K$\alpha$ feature and X-ray continuum emission whose properties depend on the accretion rate, {\it i.e.} on the density and stratification of the disc (see Fig. 8).

To study the dependence of the spectrum with the e-beam energy, an additional set of simulations has been run for 
an accretion rate of $10^{-7}$~M$_{\sun}$yr$^{-1}$ and $ <K> = 40, 100, 200, 400, 800, 1600$~Kev. As expected, as the e-beam
mean kinetic energy rises, the Fe feature becomes stronger and the centroid of the back-scattered spectrum moves to higher energies.

\begin{figure}
\begin{tabular}{l}
\includegraphics[width=14cm]{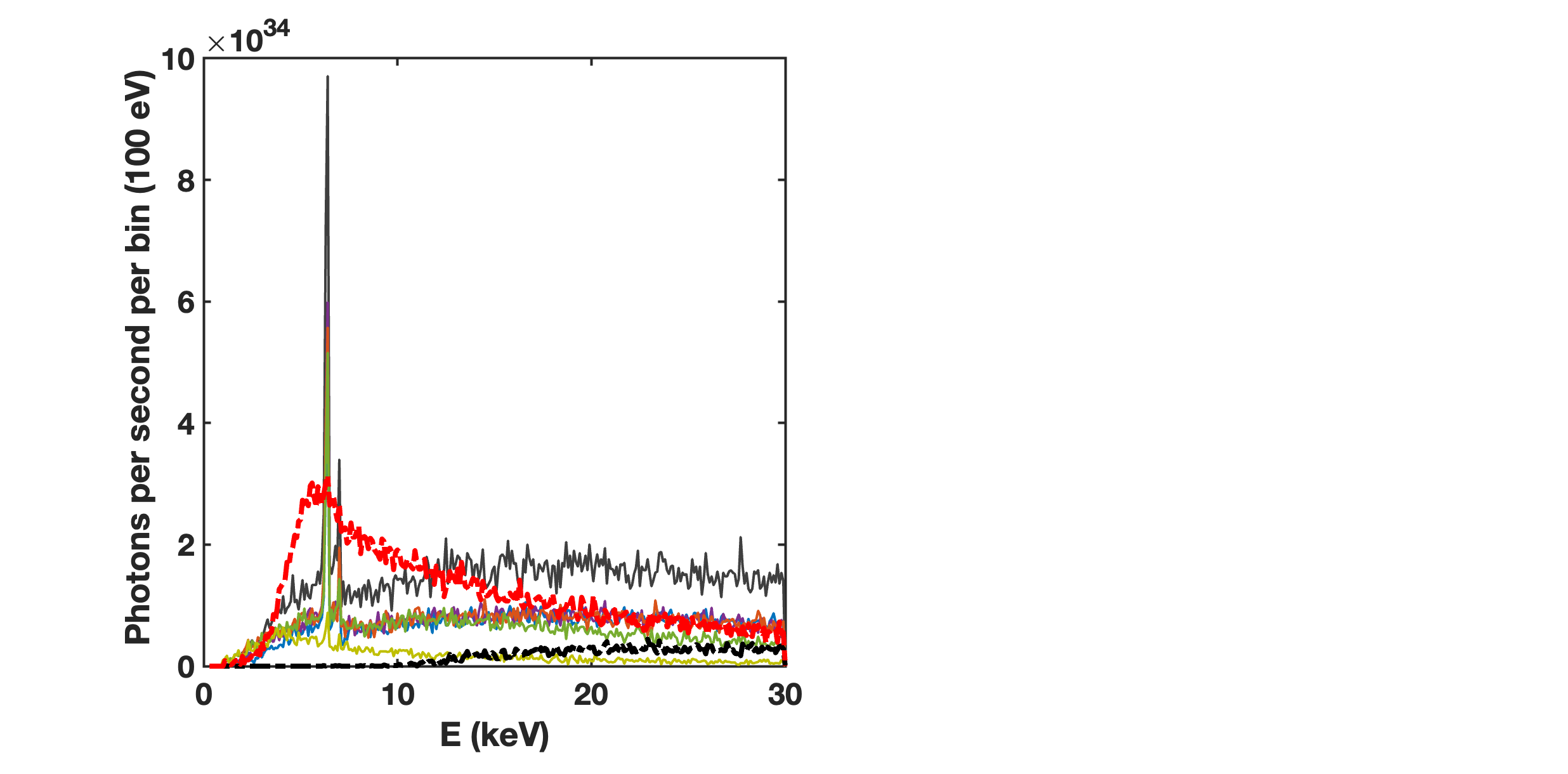}\\
\includegraphics[width=12cm]{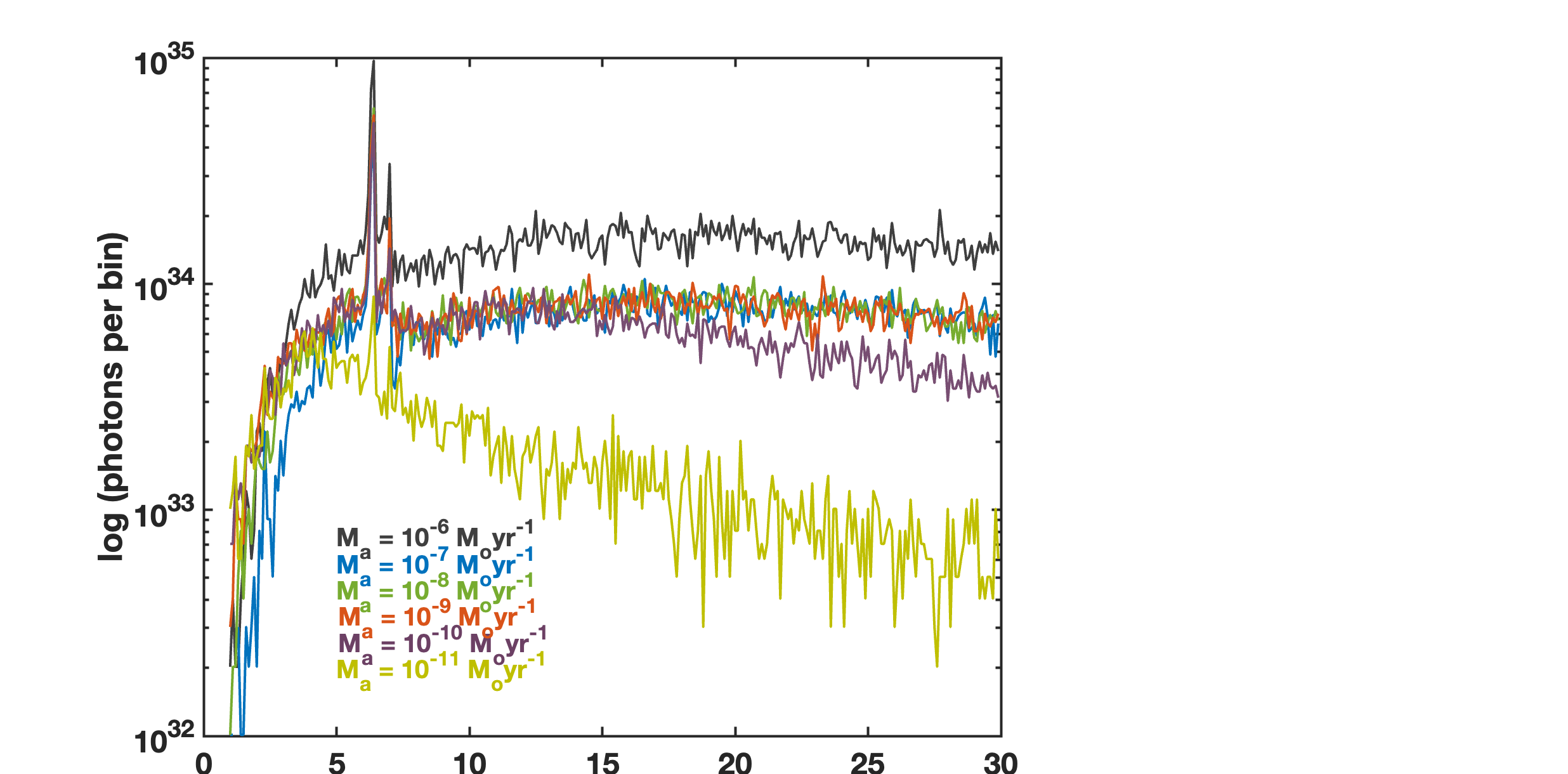}\\
\end{tabular}
\caption{Output X-ray spectrum of  the hot spot. {\it Top panel:} Comparison between the X-ray spectra of the hot spot (thin lines)
with the transmitted spectra (thick dashed lines, see Fig. 7). All the spectra are plotted in the same scale. Note the strength of the Fe feature in the back-scattered spectrum at high accretion rates. {\it Bottom panel:} X-ray spectra of the hot spot in logarithmic scale.  E-beam properties:
$K=800keV$ and   $2\times 10^{-3}$ L$_{\odot}$. All the spectra are plotted at the same scale.   }
\label{}
\end{figure}

\begin{figure}
\includegraphics[width=9.3cm]{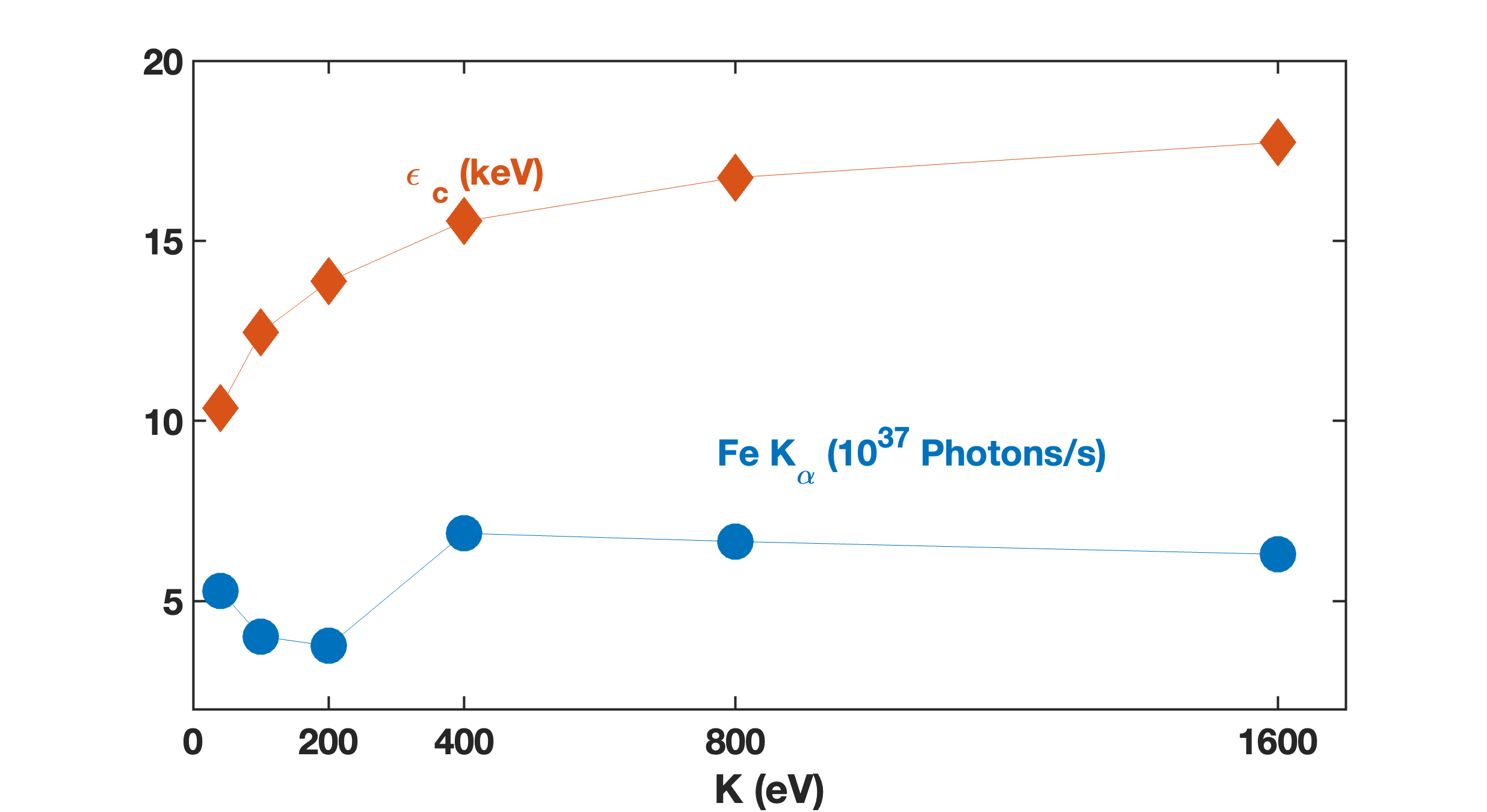} \\
\includegraphics[width=10cm]{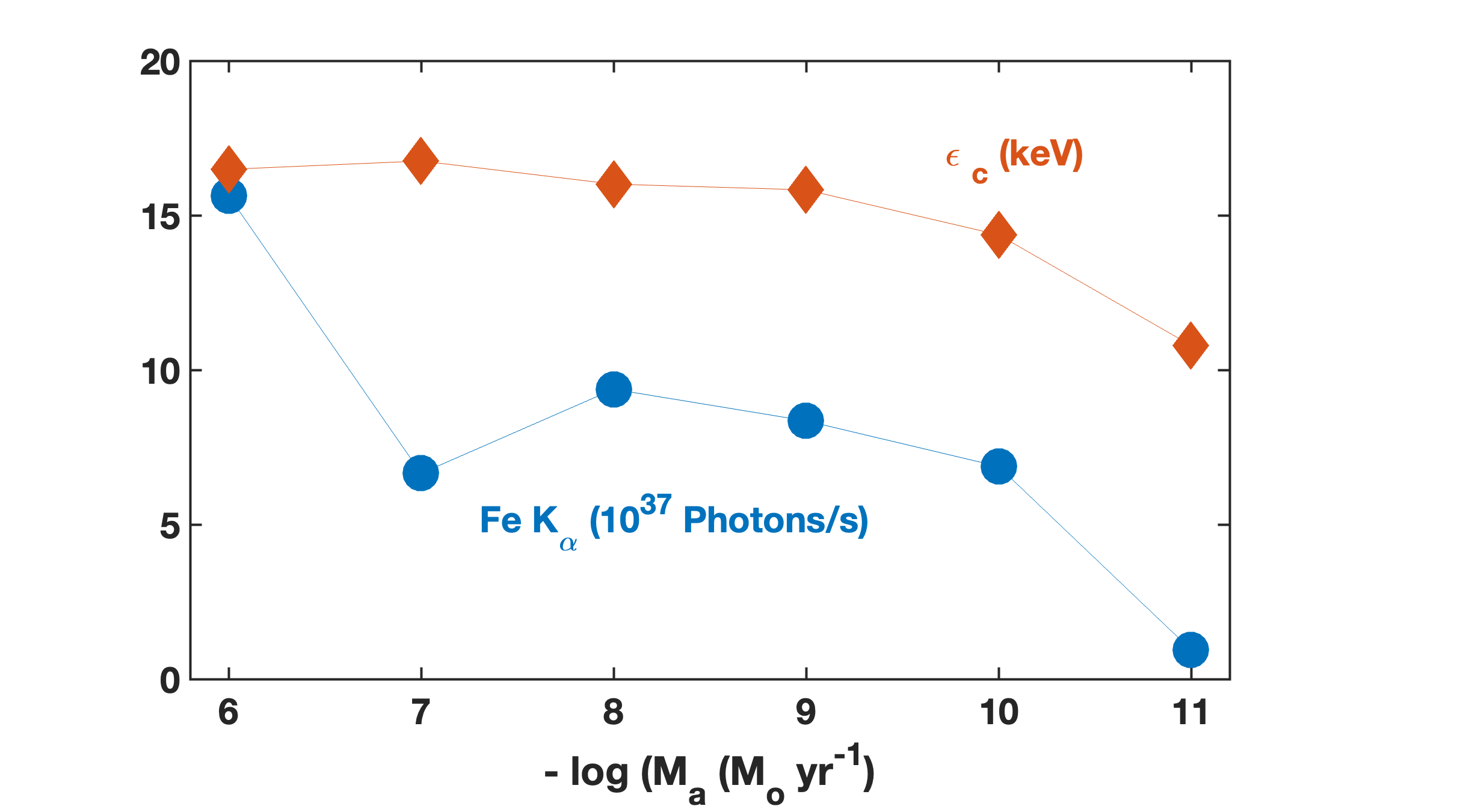}
\caption{ Variation of the centroid of the X-ray spectrum and the Fe K$\alpha$ flux with: {\it top} the average energy of the e-beam 
(for $M_a = 10^{-7}$~M$_{\sun}$yr$^{-1}$), and, {\it bottom}  the accretion rate (for K=800 keV). }
\label{}
\end{figure}

The trends described in this section can be summarized as follows,
\begin{itemize}
\item Fe K$\alpha$ (and K$\beta$) emission is very weak or absent in the transmitted spectra.
\item  Increasing the mean kinetic energy (<K>) of the e-beam or increasing the density of the medium shifts the centroid 
of the spectrum to higher energies.

\end{itemize}

Also the e-beam is highly directive, the size of the hot spot at the point of impact  is comparable to the beam radius and the 
radiative output scales linearly with the surface of spot.

\section{Interpretation of RY~Tau flares}

RY Tau is among the best studied accreting TTSs at X-rays. Its X-ray spectrum is known to vary from quiescence to flare state
from the observations carried out with the Chandra and the XMM-Newton satellites (see Skinner et al. 2016). For this work, 
we will use the XMM-Newton observations to evaluate whether the reported Fe~K$\alpha$ emission during flares can be produced from the
hot spots generated during the flares.

RY Tau was observed in September 2000 and in August 2013 (see Table 2) however, no significant flaring activity 
was detected in 2000 \citep{2007A&A...468..353G}.  We have retrieved the data obtained with the European Photon Imaging Camera (EPIC) pn instrument in 2013 from the archive,  id. 0722320101, and processed them using the XMM-Newton Science Analysis System (SAS 19.0.0).
Spectrum analysis was done using XSPEC 12.11.1.   

 \begin{table}
\caption{XMM-Newton EPIC observations of RY Tau}           
\label{table:2}    
\begin{flushleft}               
\begin{tabular}{l c c}      
\hline\hline                 
Observation ID. & Start date of observation& Duration\\
 & dd - mm- yyyy hh:mm:ss & (s) \\
\hline   
0101440701 & 05-09-2000 02:08:06 & 51,315 \\
0722320101 & 21-08-2013 04:16:33 & 112,600 \\
\hline                                   
\end{tabular}
\end{flushleft}  
\end{table}

Following an approach similar to \citet{2016ApJ...826...84S}, the spectra have been extracted using PN cleaned and filtered for
background event files. Firstly, the event files were created by removing high energy protons intervals at the end
of the exposure and by time-filtering for covering the quiescent and flare periods. The quiescent spectrum was
produced using events from $0.0$ to $20.0$ ks and from $45.0$ to $90.0$ ks, and the flare spectrum using events from
$20.0$ to $45.0$ ks. Then, source and background spectra were extracted using the {\it evselect} task. After that, the
source spectra were obtained using a computed redistribution matrix and the proper ancillary file. Finally, the spectra
were rebined using the {\it specgroup} task. The quiescent spectrum was rebinned in order to have at least $40$ counts
for each background-subtracted spectral channel. The flare spectrum was then rebinned to group it in exactly the
same way as the source spectrum. They are plotted in Figure 10 and are similar to those in Fig. 3 of  \citet{2016ApJ...826...84S}. 
 
\begin{figure}
\includegraphics[width=9cm]{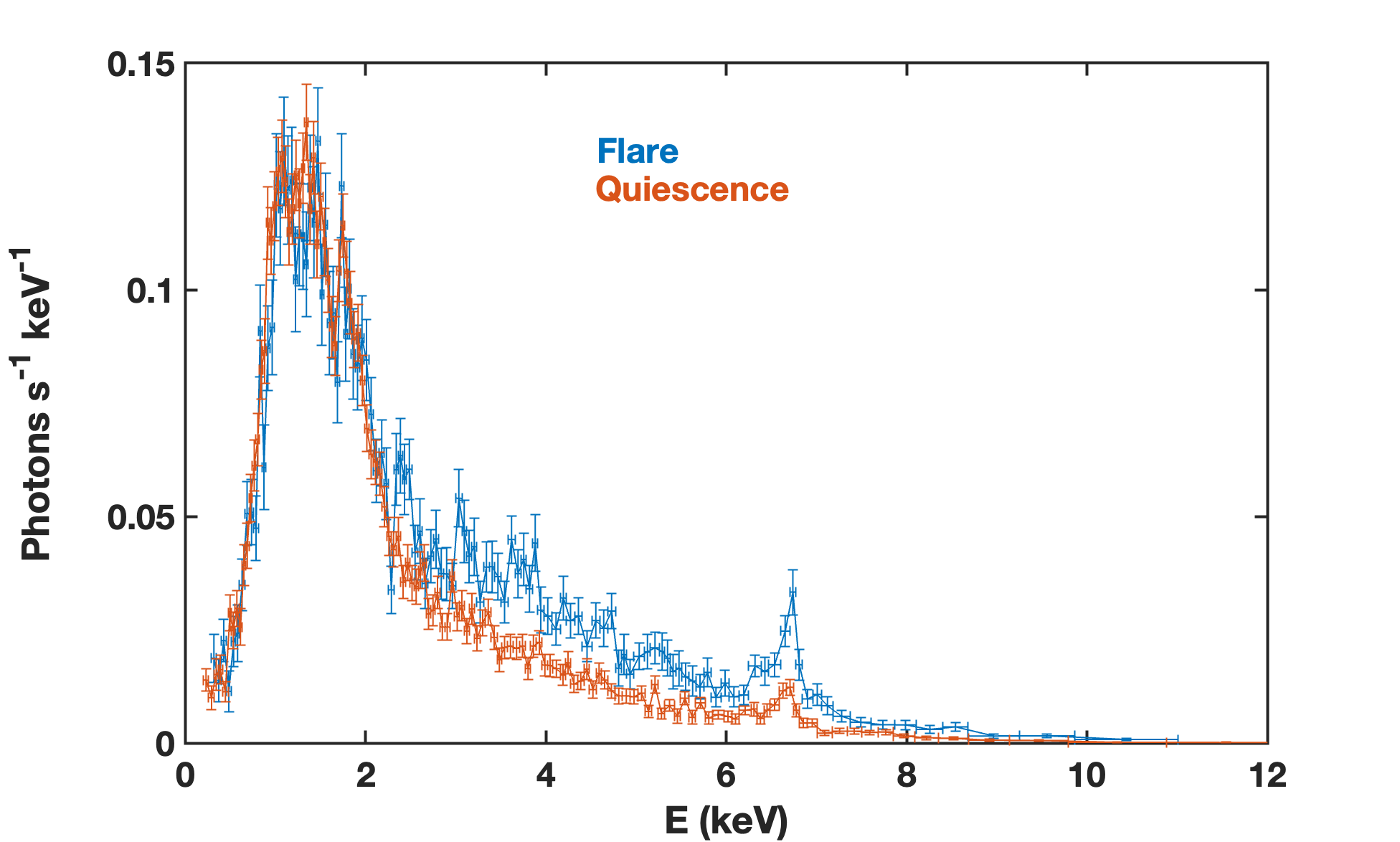}\\
\caption{X-ray spectrum of RY Tau obtained with the instrument EPIC on board the XMM-Newton satellite in August, 2013 
(observation ID: 0722320101). The X-ray light curve has been analyzed to discriminate the flare spectrum from the quiescence one.}
\label{}
\end{figure}

During the flare, the X-ray flux increases at energies above 2 keV; the prominent Fe K$\alpha$ emission
is also, clearly apparent as well as the Fe XXV feature. The spectral energy distribution of the  X-ray 
emission during the flare is represented in Fig. 11. To obtain it, the quiescence 
spectrum is subtracted out from the flare spectrum but only energy bins with SNR$\geq 3$ 
have been considered. The error bars have been computed using the standard expression for the linear propagation of errors:
\begin{small}
\begin{equation}
\sigma _{excess}^2 = \sigma _{flare}^2+\sigma _{quiescence}^2
\end{equation}
\end{small}
\noindent
with $\sigma_{flare}$ and $\sigma_{quiescence}$ the variances of the flare and quiescence spectra provided by XSPEC
(for obvious reasons, some data points have $\sigma _{excess} < 3$).

\begin{figure}
\includegraphics[width=9cm]{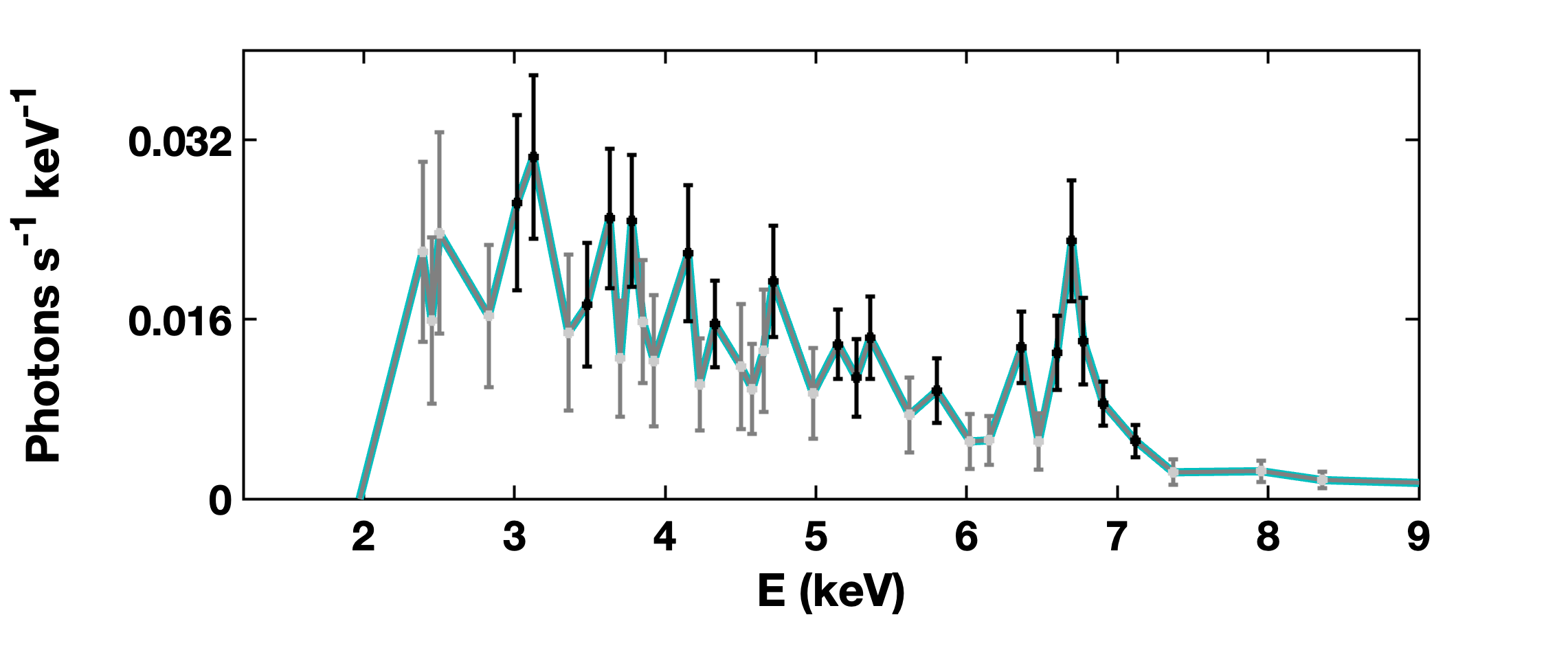}\\
\caption{Spectral distribution of the X-ray radiation produced by the flare. Data points with SNR$\geq 3$ are marked in black and
those with SNR$\geq 2$ in grey. The green line is plotted to guide the eye to the overall energy distribution where the prominent
Fe~K$\alpha$ and Fe XXV features are readily recognized.}
\label{}
\end{figure}

The Fe~K$\alpha$ flux during the flare has been computed using the XPEC routines
and found to be: $0.87 \times 10^{-13}$~erg~s$^{-1}$~cm$^{-2}$, which accounts to a total luminosity of the line of
$2.01 \times 10^{29}$~erg~s$^{-1}$, using the {\it Gaia} parallax of RY~Tau (7.2349 mas) for the calculation. 
This luminosity is equivalent to a grand total of $1.96 \times 10^{37}$ Fe~K$\alpha$ photons per second during the flare. 
This value should be divided by $4\pi$ to compare it with the results from the simulations described in the previous sections.

This luminosity cannot be accounted by  the impact of relativistic electrons on the diffuse environment around the loops 
(either the funnel flows or the dense stellar corona) since, as shown in Sect.~3, Fe~K$\alpha$ radiation is only produced if 
the density is very high ($>10^{14}$~cm$^{-3}$). However, the observed Fe~K$\alpha$ emission could arise from hot spots on the accretion 
disc generated at the impact point of the e-beams from the flares. 

The environment  around an active TTS such as RY Tau is strongly magnetized and has a complex topology prone to
produce magnetic reconnection in many different regions. A simplified approach to  the overall topology is displayed in Fig.~12  \citep[after][]{2016JATIS...2d1215G}; the magnetic field is represented by solid lines with the very thick line marking the connection between the 
star and the inner border of the disc. In the classical TTS evolutionary state, the stellar field is connected with the disc field to the point that stellar rotation is locked to the Keplerian velocity of the inner border of the disc. The dashed lines mark the limit between the disc-dominated and 
the star-dominated regions. Over this magnetic sketch, the results of numerical simulations dealing with this interaction  \citep{2011MNRAS.411..849G} are shown in the figure; the green-brown shadowed regions are locations where magnetic
reconnection occurs heating the gas and producing ultraviolet radiation.  In this environment, the most obvious region to have 
frequent and strong reconnection events is the inner border of the disc. However, the geometry is not clear; reconnection may 
occur above the disc producing e-beams impacting on the disc surface (as simulated in Sect. 4) however, e-beams may be 
generated just at the inner border and interact with specific layers of the disc vertical structure. 
These two simple possibilities are marked in Fig.~12 as Option A and B, respectively.
Though the geometry is going to be much more complex in general, let us keep simple for this prospective work 
and evaluate whether the Fe~K$\alpha$ emission detected during the flare is compatible with the 
radiation produced in the propagation of the high energy e-beams produced during large flares within the disc.

\begin{figure*}
\includegraphics[width=16cm]{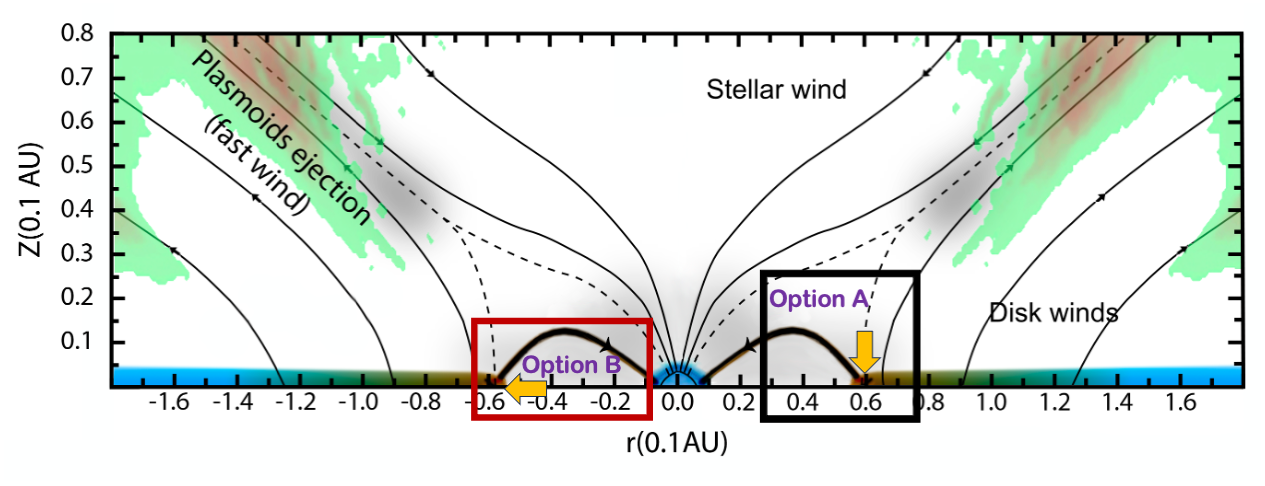} \\
\caption{ Map of the C III] (191 nm) emissivity caused by the star-disc interaction from MHD simulations of the disc-star interaction \citep{2011MNRAS.411..849G}. On top, the main components of the TTSs environment and magnetic structure; stellar interaction with the infalling magnetized plasma drives the outflow and generates reconnection layers (dashed lines). }
\label{}
\end{figure*}

According to the simulations in Sect. 4, the Fe~K$\alpha$ luminosity produced by an e-beam impacting perpendicularly to the 
accretion disc depends on the accretion rate which has been measured to range between $0.91 \times 10 ^{-7}$~M$_{\odot}$~yr$^{-1}$ 
and $0.50 \times 10 ^{-7}$~M$_{\odot}$~yr$^{-1}$ for RY Tau \citep{2021A&A...652A..72A}. For these rates, the predicted line luminosity ranges between $3.4\times 10 ^{37}$~photons~s$^{-1}$ and $4.7\times 10 ^{37}$~photons~s$^{-1}$ for a mechanical luminosity of the e-beam of 
0.1~L$_{\odot}$, which is within a factor of two of the detected values. This factor could be easily accommodated reducing the luminosity of the flare
or the surface of the hot spot.

In the case of a lateral impact (option B in Fig 12), the particle density of the mid-plane  is expected to range between $0.77 \times 10^{16}$~cm$^{-3}$ and $1.13 \times 10^{16}$~cm$^{-3}$ (see Eq. 8) and thus, according to Fig.~5, the Fe~$K\alpha$ emission 
is predicted to range between $3 \times 10^{36}$~photons~s$^{-1}$  and $6 \times 10^{36}$~photons~s$^{-1}$ depending of the mean
kinetic energy of the electrons, for an e-beam of mechanical  luminosity of 0.02~L$_{\odot}$. Since, the  Fe~$K\alpha$ luminosity scales 
linearly with the mechanical luminosity, a strong flare with a luminosity 0.1~L$_{\odot}$ could produce an emission similar to the
observed.

In summary, the Fe~K$\alpha$ emission detected in RY Tau during the large flare in 2013 can be easily accounted by the
propagation of the e-beams generated during the flare within the inner border of the disc.

\section{Conclusions}

The  X-ray observations of the TTSs have shown that some classical TTSs  experience very strong flares which are associated
with the release of magnetic energy during the reconnection of very long magnetic loops, extending  few stellar radii. This strong reconnection 
events are possibly produced when stellar loops meet the inner of the disc. A characteristic of these flares is the detection of Fe~K$\alpha$
emission which, in the case of RY~Tau, reaches a line luminosity of $1.96 \times 10^{37}$ Fe~K$\alpha$ photons~s$^{-1}$ during the flare. 
In this work, we have shown that this emission can be naturally explained by the X-ray radiation produced at the impact point of 
the high energy electrons released during the reconnection event on the accretion disc.

This works also includes a broad network of numerical simulations of the expected X-ray radiative output in this context to
aid the interpretation of  X-ray observations of TTSs during flares. 

This work has concentrated in modelling the radiative output from large flares meeting the inner border of the disc but similar phenomena 
(and Fe~K$\alpha$ emission) could be produced by the impact of the e-beams on the stellar surface however, in all cases, 
very strong flare luminosities are required to reproduce the observed Fe~K$\alpha$ flux.

\begin{acknowledgements}
This work has been partially funded by the Ministry of Science, Innovation and Universities of Spain through the grant with reference: PID2020-116726RB-I00.

\end{acknowledgements}

\bibliographystyle{aa}
\bibliography{references.bib}

\end{document}